\documentclass{article}

\usepackage[utf8]{inputenc} 
\usepackage[T1]{fontenc}    
\usepackage{hyperref}       
\usepackage{url}            
\usepackage{booktabs}       
\usepackage{amsfonts}       
\usepackage{nicefrac}       
\usepackage{microtype}      
\usepackage{lipsum}
\usepackage{rotating}
\usepackage{threeparttable}
\usepackage{amsmath,amssymb}
\usepackage{caption}
\begin{document}
\baselineskip=24pt
\begin{center}
\Large 
Using clinical trial registries to inform Copas selection model for publication bias in meta-analysis
\end{center}

\vspace{3mm}

\begin{center}
Ao Huang ,
Sho Komukai,
Tim Friede ,
Satoshi Hattori\renewcommand{\thefootnote}{\fnsymbol{footnote}}
\footnote[1]
{\textit{correspondence to:} Satoshi Hattori,Institute for Open and Transdisciplinary Research Initiatives, Osaka University; email: \textbf{hattoris@biostat.med.osaka-u.ac.jp}}
\end{center}

\begin{abstract}
Prospective registration of study protocols in clinical trial registries is a useful way to minimize the risk of publication bias in meta-analysis, and several clinical trial registries are available nowadays. However, they are mainly used as a tool for searching studies and information submitted to the registries has not been utilized as efficiently as it could. 
In addressing publication bias in meta-analyses, sensitivity analysis with the Copas selection model is a more objective alternative to widely-used graphical methods such as the funnel-plot and the trim-and-fill method. Despite its ability to quantify the potential impact of publication bias, 
a drawback of the model is that some parameters not to be specified. This may result in some difficulty in interpreting the results of the sensitivity analysis. In this paper, we propose an alternative inference procedure for the Copas selection model by utilizing information from clinical trial registries. Our method provides a simple and accurate way to estimate all unknown parameters in the Copas selection model. A simulation study revealed that our proposed method resulted in smaller biases and more accurate confidence intervals than existing methods. Furthermore, two published meta-analyses had been re-analysed to demonstrate how to implement the proposed method in practice.

\end{abstract}

\baselineskip=15pt

\section{Introduction}\label{sec1}

Publication bias is one of the most  important concerns in systematic reviews and meta-analyses. Part of the issue is that authors and scientific journals are likely to publish studies with statistically significant results than inconclusive studies\cite{thornton2000}. To be statistically significant, the studies with small sample size need to show larger treatment effects, which is known as the small-study effect\cite{sterne2000}.  Thus, this may force us to analyze selective samples from the population of interest, and the standard meta-analysis techniques may suffer from this selection bias, which is referred to as publication bias.

The funnel plot is a simple and widely-used graphical tool to check for publication bias, which is defined by plotting the estimate of the effect size such as the log-odds ratio or the log-hazard ratio as the horizontal axis and the related measure of precision such as the square root of the sample size or the inverse of the standard error as the vertical axis. One can address the publication bias by examining asymmetry of the funnel plot visually and more formal statistical evaluation of the funnel-plot asymmetry can be made with a regression based test\cite{egger1997,macaskill2001} or a rank-based test \cite{begg1994, schwarzer2007}. The trim-and-fill method is a nonparametric method to adjust publication bias utilizing the funnel-plot asymmetry\cite{duval2000}. 

All these methods are simple to apply and are widely used in practice. However, publication bias is not the only reason for asymmetry of the funnel plot; Egger et al.\cite{egger1997} listed different potential reasons for the funnel-plot asymmetry including the between-study heterogeneity and others. Then, detection of the publication bias by the funnel plot may be subjective and the trim-and-fill method may not work well\cite{terrin2003, peters2007}. Alternatively, several sensitivity analysis methods have been developed. Copas and colleagues introduced the Heckman-type parametric selection model, which was originally proposed in the econometrics literature\cite{heckman1979}, to describe a selective publication process in meta-analysis. In this paper, this selection model is referred as the Copas selection model\cite{copas2000,copas2001}. The sensitivity analysis with the Copas selection model has several advantages over existing alternative sensitivity analysis methods\cite{copas2013,copas2004,henmi2007}; a user-friendly \emph{R} package \emph{metasens} is available for the Copas-Shi selection model\cite{copas2000,copas2001}, and its performance was empirically examined\cite{carpenter2009, schwarzer2010}. Thus, we focus on this selection model in this paper.    

The Copas selection model consists of a pair of models; one is for the outcome, such as the log-odds ratio or the log-hazard ratio reported in publications, and the other is a model for the latent variable describing a selective publication process. It's hard to estimate all the parameters in the Copas selection model, and thus a series of papers by Copas and colleagues took a sensitivity analysis approach; some parameters were fixed as sensitivity parameters in a certain range, and the impact on the estimate of interest was studied. However, it's not easy to define an appropriate range for the sensitivity parameters.

Prospective registration of study protocols in clinical trial registries is a useful non-statistical approach for publication bias. The International Committee of Medical Journal Editors (ICMJE) initiated a trials-registration policy as a condition for publication in its member journals in July 2005\cite{deangelis2005}, which promoted the development and utilization of the clinical trial registries. ClinicalTrials.gov (\url{https://clinicaltrials.gov/ct2/home}) is one of the early examples which was established in 2000 and is nowadays a very popular registry with 334,954 registered studies in April 2019. World Health Organization's (WHO) International 
Clinical Trials Registry Platform (ICTRP) (\url{http://apps.who.int/trialsearch/}) is another well-known clinical trials registry which was constructed after the initiation of ICMJE's policy and has been supported by the WHO. FDA (US Food and Drug Administration) database
 (\url{https://www.accessdata.fda.gov/scripts/cder/daf/}) contains information about products approved by the FDA for human use in the United States. Besides, EU clinical trials register (\url{https://www.clinicaltrialsregister.eu/ctr-search/search}) and ISRCTN (\url{https://www.isrctn.com/}) are also well-developed and widely used in recent years as well as many national registries that often also provide lay summaries in the local language\cite{viergever2015,ogino2014}. 

Baudard et al. \cite{baudard2017} investigated 14 meta-analyses and showed that by incorporating the information from clinical trial registries, the effect size estimates might change by 29\%. It indicated that searching clinical trial registries may be useful to reduce the bias due to selective publication. On the other hand, although clinical trial registries provide potentially useful information of the registered clinical trials (e.g., study design, target sample size, development phase), they are used only as a tool to search for studies\cite{jones2014}. Information on unpublished studies provided by clinical trial registries is not formally incorporated in the analyses. To the best of knowledge, only two papers mentioned the use of clinical trial registries in statistical inference. Matsuoka et al. \cite{matsuoka2011} proposed a publication bias detection method by utilizing the planned  sample size from World Health Organization's (WHO) International Clinical Trials Registry Platform (ICTRP). Mavridis et al. \cite{mavridis2013}utilized FDA database to construct a prior distribution of parameters in their Bayesian Copas-type model for sensitivity analysis in network meta-analysis. 

In this paper, we improve statistical inference for the Copas-Shi selection model utilizing the information from clinical trial registries. To be specific, utilizing the planned sample size of unpublished studies, which are provided by most clinical trial registries, we propose a method that fit by the maximum likelihood method. All the unknown parameters can be estimated from data, and statistical inference can be conducted according to the standard maximum likelihood theory. In addition to the confidence interval with the normal quantile according to the maximum likelihood theory, we also propose two modified confidence intervals aiming improvement of performance with the small number of studies. Through numerical studies, we observed that our method outperformed existing methods with smaller biases and coverage probabilities closer to the nominal level. Thus, by utilizing the information of unpublished studies from clinical trial registries, inference based on the Copas selection model can be improved.

The remainder of this paper is organized as follows. In Section 2, we introduce two motivating meta-analyses. In Section 3, we briefly review the sensitivity analysis method of Copas and Shi\cite{copas2000,copas2001}, and then introduce our new inference procedure by utilizing clinical trial registries. In Section 4, we conduct a series of simulation studies motivated by the tiotropium study to evaluate the performance of our proposed method in comparison to other competitive approaches. In Section 5, we revisit the two studies in Section 2 to illustrate the proposed method. Finally, we discuss some limitations and further research in Section 6.

\section{Motivating examples}\label{sec2}

\subsection{Tiotropium study}
Karner et al.\cite{karner2014}conducted a Cochrane review of tiotropium versus placebo in COPD. The primary outcome was COPD exacerbations, which was defined as a complex of respiratory events or symptoms (new onset or an increase in at least one of cough, sputum, dyspnoea or wheeze) that lasted at least three days and required treatment with antibiotics and/or systemic corticosteroids. In each study, the odds ratio for the event of one or more exacerbations was used to compare the treatment and placebo groups. In the meta-analysis by Karner et al.\cite{karner2014}, 22 studies were analyzed using a random-effects model, and the pooled odds ratio was reported as 0.78 with a 95\% confidence interval of [0.70, 0.87]. 

To assess the potential risk of publication bias,  Karner et al.\cite{karner2014}conducted the Egger's test with the p-value of 0.22 and concluded no serious concern of publication bias. However, our additional application of the Macaskill's regression test\cite{macaskill2001} gave a p-value of 0.06, suggesting the funnel-plot asymmetry. Then, publication bias might be a concern. Karner et al. \cite{karner2014}made their efforts to search the studies with results submitted on ClinicalTrials.gov. We conducted a more comprehensive search of multiple registries, including ClinicalTrials.gov, World Health Organization's (WHO) International Clinical Trial Registry, EU clinical trials register, FDA database, and ISRCTN platform. We used the following terms: \emph{Tiotropium} or \emph{Spiriva} or \emph{HandiHaler} or \emph{Respimat} for the search. They were the same as used in the meta-analysis of Karner et al. \cite{karner2014}, and we also set  primary completion date  before February 28, 2012, which aimed to be consistent with the original meta-analysis. We detected 10 related studies on these registries, among which 8 came from ClinicalTrials.gov, 1 came from EU clinical trials register, and 1 came from World Health Organization's (WHO) International Clinical Trials Registry. Two of them reported the outcome measure and were incorporated with published studies,  and 8 studies without results were deemed as unpublished studies. Finally, we got the tiotropium dataset with 24 published studies and 8 unpublished studies with information of sample size only (see Table S1 in Web-appendix A). 

\subsection{Clopidogrel study}
The second example is about the comparison of high and standard maintenance-dose clopidogrel on major adverse cardiovascular/cerebrovascular events (MACE/ MACCE) by Chen et al.\cite{chen2013}. The original meta-analysis included 12 studies and reported pooled odds ratio as 0.6 based on the fixed-effect model with a 95\% CI of  [0.43, 0.83], and no significant bias was concluded with the Egger's test (p=0.25). However, we rerun the  Macaskill's test, which suggested funnel-plot asymmetry with a P value of 0.02. We took the procedure as given in the first example and identified 3 unpublished studies from ClinicalTrials.gov by following terms: \emph{Clopidogrel} or \emph{Plavix} or \emph{Iscover}, with completion date  before August 31, 2013. The resulting dataset was presented in Table S2 in Web-appendix A.
 
\section{Inference procedures for the Copas selection model}\label{sec3}

\subsection{Brief review of the sensitivity analysis method by Copas and Shi}

In order to evaluate the potential impact of publication bias on estimation of the treatment effect, Copas and Shi \cite{copas2000} introduced a sensitivity analysis method. Suppose we conduct a meta-analysis of \emph{N} studies. Let $\hat{\theta_i}$ be the estimated treatment effect of the $i$th study such as the log-odds ratio or the log-hazard ratio, and $s_i$ be an estimate of its standard error. We assume that $(\hat{\theta_i}, s_i)$ is available for $i=1,2,..., N$.
To integrate the treatment effects over the studies, we consider the standard random-effects model with
\begin{eqnarray}
\hat{\theta_i}=\theta_i+\sigma_i\epsilon_i, \ \ \ \theta_i \sim N(\theta,\tau^2), \ \ \ \ \epsilon_i \sim N(0,1),
\label{mixed}
\end{eqnarray}
where $\theta_i$ is the study-specific treatment effect of the \emph{i}th study, which is a random-effect following a normal distribution with mean $\theta$ and between-study variance $\tau^2$, and $\sigma_i^2$ is the true within-study variance of the outcome $\hat {\theta_i} $. The random elements $\theta_i$ and $\epsilon_i$ are assumed to be independent. 

To describe the selective publication process, Copas and Shi \cite{copas2000,copas2001} considered the following model
\begin{equation}
Y_i=\alpha_0+\alpha_1 / s_i+\delta_i, \ \ \ \delta_i \sim N(0,1), \ \ \ corr(\epsilon_i,\delta_i)=\rho,
\label{selection}
\end{equation}
where $corr(\epsilon_i,\delta_i)$ denotes the correlation coefficient between $\epsilon_i$ and $\delta_i$. The \emph{i}th study is assumed to be published if and only if $Y_i>0$.  The term $\alpha_0+\alpha_1 / s_i$ models the small study effect. That is, studies of high precision are more likely to be pulished. The correlation coefficient $\rho$ is responsible for describing the influence of the outcome $\hat{\theta_i}$ on likelihood of publication. Copas and Shi\cite{copas2000} considered to make inference by maximizing the likelihood function conditional on the study being published. The conditional likelihood function is given by
\begin{eqnarray}
&& L_{obs}(\theta,\tau,\rho,\alpha_0,\alpha_1;s_i)=\sum_{i=1}^{N}\left[log f(\hat{\theta_i}|Y_i>0,s_i)\right] \nonumber \\                                                                     
&&= \sum_{i=1}^{N}\left[-\frac{1}{2}log(\tau^2+\sigma_i^2)-\frac{(\hat{\theta_i}-\theta)^2}{2(\tau^2+\sigma_i^2)}-log\Phi(\alpha_0+\alpha_1/s_i)+log\Phi(v_i)\right], 
\label{lik_obs}
\end{eqnarray}  
where $f(\cdot|Y_i>0,s_i)$ is the probability density function of $\hat{\theta_i}$ conditional on $Y_i>0$ and $s_i$. $ \Phi(\cdot)$ is the cumulative distribution function of the standard normal distribution, and $v_i=\left\{\alpha_0+\alpha_1/s_i +\rho \sigma_i(\hat{\theta_i}-\theta) / (\tau^2+\sigma_i^2) \right\} / \sqrt {1-\rho^2 \sigma_i^2 / (\tau^2+ \sigma_i^2)}$.  In meta-analysis literatures, $s_i$ is usually regarded as being equal to $\sigma_i$ and known. This is relevant approximately at least when the number of subjects is sufficiently large. Copas and Shi\cite{copas2000,copas2001}took care of underestimation of $\sigma_i$ due to publication bias, and then they utilized  the quantity $var(\theta_i|s_i,Y_i>0)$ to estimate $\sigma_i^2$ in inference of their selection model. As argued by Copas and Shi\cite{copas2000,copas2001}, the log-likelihood function ($\ref{lik_obs}$) may take its maximum over a very  flat plateau, and then it may be  computationally very challenging to estimate all the parameters simultaneously. Thus Copas and Shi\cite{copas2000,copas2001} proposed that the vector $(\alpha_0,\alpha_1)$ was fixed as sensitivity parameters, and $(\theta,\tau,\rho)$ was estimated  by maximizing the conditional log-likelihood function ($\ref{lik_obs}$). One can examine how the estimate $\hat{\theta}$ for the overall treatment effect $\theta$ would be influenced by the assumed selective publication process. Since we do not know the true value of $(\alpha_0,\alpha_1)$, we should apply this method with various choices for the parameter vector $(\alpha_0,\alpha_1)$ in a certain range. According to the formula by Copas and Shi\cite{copas2000}, one can translate $(\alpha_0,\alpha_1)$ into the expected numbers of unpublished studies by $ M=\sum_{i=1}^{N}\left\{1-P(Y_i>0|s_i)\right\}/P(Y_i>0|s_i)$, which is more interpretable. Then, one can evaluate how biased the estimate of the treatment effect $\theta$ is with a certain number of unpublished studies behind. Although such a consideration provides nice insights about robustness of meta-analysis results, it is still unclear what range of the number of unpublished studies is sufficient to be considered.

\subsection{Proposed inference procedure utilizing clinical trial registries}

In addition to the \emph{N} published studies, we suppose that there are \emph{M} unpublished studies identified by searching clinical trial registries. Without loss of generality, the first \emph{N} studies are published and the last \emph{M} studies are unpublished. Items required at registration are not necessarily common among clinical trial registries. For example, FDA database provides the sample size of each experimental group in its summary review PDF document, but other clinical trial registries may not give it. However, all the clinical trial registries we referred provide the planned total sample size (not separately by the experimental groups) of each study. Thus, we suppose the triple $(n_i,\hat{\theta_i},s_i)$ is available for \emph{i}=1,2,...,\emph{N} (published studies), and only $n_i$ is available for \emph{i}=\emph{N}+1,..,\emph{N+M} (unpublished studies). A binary variable $D_i=I(Y_i>0)$ denotes the publication status (published/unpublished) of each study. Instead of  the model ($\ref{selection}$), we employ an alternative selection model
\begin{eqnarray}
Y_i=\alpha_0+\alpha_1\sqrt{n_i}+\delta_i, \ \ \ \delta_i \sim N(0,1),
\label{selection2}
\end{eqnarray}
which was used by Copas\cite{copas1999}.  For the models ($\ref{mixed}$) and ($\ref{selection2}$), we introduce the maximum likelihood approach as follows.

Considering the likelihood function of the published and the unpublished studies, the log likelihood is given by
\begin{align}
L_{full}(\theta,\tau,\rho,\alpha_0,\alpha_1)&=\sum_{i=1}^{N}\left[log f(\hat{\theta_i}|Y_i>0;n_i)+log P(Y_i>0;n_i)\right]\nonumber \\&+\sum_{i=N+1}^{N+M}\left[log P(Y_i<0;n_i)\right]\nonumber  \\
&=\sum_{i=1}^{N}\left[-\frac{1}{2}log(\tau^2+\sigma_i^2)-\frac{(\hat{\theta_i}-\theta)^2}{2(\tau^2+\sigma_i^2)}+log\Phi(\tilde{v_i})\right]\nonumber \\&+\sum_{i=N+1}^{N+M}\left\{log [1-\Phi(\alpha_0+\alpha_1\sqrt{n_i})]\right\},
\label{lik_full}   
\end{align}
where  $\tilde{v_i}=\left\{\alpha_0+\alpha_1\sqrt{n_i} +\rho \sigma_i(\hat{\theta_i}-\theta) / (\tau^2+\sigma_i^2) \right\} / \sqrt {1-\rho^2 \sigma_i^2 / (\tau^2+ \sigma_i^2)}$. 
Following the treatment often made in meta-analysis literature, we suppose that $\sigma_i=s_i$. By maximizing  this log-likelihood function ($\ref{lik_full}$), we can estimate all the parameters $(\theta,\tau,\rho,\alpha_0,\alpha_1)$ simultaneously, and the resulting maximum likelihood estimator is denoted by $(\hat{\theta},\hat{\tau},\hat{\rho},\hat{\alpha_0},\hat{\alpha_1})$. Maximization of the log-likelihood function can be implemented with the standard non-linear optimization techniques. Here, we employ the constraint optimizer L-BFGS-B method in  the nlminb() function  in \emph{R}  (package stats, version 3.6.2) to do this. Statistical inference can be conducted following the standard maximum likelihood theory.

\section{Simulation study}\label{sec4}

\subsection{Methods to be compared}

Simulation studies were carried out to evaluate the performance of our new inference procedure. We investigated whether or not the proposed method outperforms the Copas sensitivity analysis method. For reference, we also compared it with standard random-effects meta-analysis using the restricted maximum likelihood (REML) estimation, which did  not account for selective publication process. In addition to the standard REML confidence interval based on normal quantiles, the Hartung and Knapp method\cite{hartung2001,hartung2001refined} was applied using \emph{t}-quantiles and rescaled standard errors, which is referred to as REML.KnHa in the following.  As outlined in Section 3, the Copas method is a sensitivity analysis approach resulting in  multiple estimates under several settings of the number of unpublished studies. 
For comparison with other methods, we took the estimate of the treatment effect with the smallest number of unpublished studies of the p-value of the goodness-of-fit test larger than 0.1\cite{copas2000}, which is presented in the output of \emph{Copas} function in \emph{metasens} package\cite{carpenter2009r}.  For the REML, we used the commonly used \emph{metafor} package with method option equal to ``REML'' and test equal to ``knha'' to obtain the CI with the Hartung and Knapp method.

\subsection{Data generation}

We mimicked meta-analyses motivated by the tiotropium study. Suppose each study aimed to compare two groups of the treatment group and the control group with respect to a binary outcome. The log-odds ratio was used as the summary measure of the treatment effect between two groups. Let $\hat{\theta_i}$ in the model ($\ref{mixed}$) denoted the empirical log-odds ratio of the $i$th study. We set $\theta=-0.25$ and $\tau$=0.05, 0.15 or 0.3. The total number of studies including published and unpublished was set as 15, 25, 50 or 100. We generated datasets according to the models ($\ref{mixed}$) and ($\ref{selection2}$) as follows. At first, we generated $\theta_i \sim N(\theta,\tau^2)$, which was the true log-odds ratio of the \emph{i}th study. Next, we generated individual participant data of the two groups with the log-odds ratio of $\theta_i$. We set the true event rate of the control group from the uniform distribution \emph{U(0.2,0.9)}, and then set the event rate of treatment group satisfying the log-odds ratio of $\theta_i$. Similarly to Kuss\cite{kuss2015}, the total sample size of each study was generated from \emph{LN(5,1)}, the log-normal distribution with the location parameter 5 and scale parameter 1, and the minimum sample size was restricted to 20 patients (values below 20 were rounded up to 20). Subjects were allocated to the two groups with probability of 0.5. With the generated individual participant data, we calculated an empirical log odds ratio $\hat{\theta_i}$ and its standard error $s_i$. $Y_i$ in the model ($\ref{selection2}$) was generated according to the conditional distribution,
\begin{equation}
Y_i|\hat{\theta_i}\sim N(\alpha_0+\alpha_1 \sqrt n_i+\rho\sigma_i (\hat{\theta_i}-\theta)/(\tau^2+\sigma_i^2),1-\rho^2\sigma_i^2/(\tau^2+\sigma_i^2))
\end{equation} and set $D_i=I(Y_i>0)$, which entailed us to generate $(\hat{\theta_i},Y_i)$ from the joint distribution defined by the models ($\ref{mixed}$) and ($\ref{selection2}$). The parameters $(\alpha_0,\alpha_1)$ were set based on our consideration for the publication rate of a study with the minimum sample size of 20 and that with a large sample size of 500, which were denoted by $P_{20}$ and $P_{500}$, respectively. Set $P_{500}=0.99$, reflecting our belief that a study with 500 patients had sufficiently large probability to be published, and $P_{20}$ was set to 0.1, 0.3 or 0.5, which represented our different concern to the publication rate of a small study with sample size of 20. Thus according to the model (4), we could derive the parameters $(\alpha_0,\alpha_1)$ by solving the equations $P_{20}=\Phi(\alpha_0+\alpha_1\sqrt20)$ and  $P_{500}=\Phi(\alpha_0+\alpha_1\sqrt500)$.  Then we got three pairs of $(\alpha_0,\alpha_1)$ as (-2.18,0.20), (-1.24,0.16) and (-0.58,0.13), which resulted in 40\%, 27\% and 19\% unpublished studies on average in the simulated meta-analyses. The parameter of $\rho$ was set to -0.4 or -0.8, and for each $\rho$ we investigated 3 scenarios of $\tau$ (0.05, 0.15 or 0.3). For each scenario, we generated 1000 simulated meta-analyses.

In Figure 1, we presented funnel plots of randomly selected simulated meta-analyses under different settings of $(\rho,\tau)$ when the total number of published and unpublished studies was 50. The filled and open circles represent published and unpublished studies, respectively, and we observed our simulation setting successfully simulated the selective publication process often observed in practice. The bold vertical line represents the true $\theta$ and the dashed vertical line is the REML estimate with published studies. They were certainly different and thus publication bias had certain impacts in simulated data.

\subsection{Simulation results}
The properties of each method were assessed by evaluating biases and standard errors of the estimates for $\theta$ and average lengths and coverage probabilities of two-tailed $95\%$ confidence intervals of $\theta$. Following the standard maximum likelihood theory, we constructed the two-tailed 95\% confidence interval for the estimates of  $\theta$ with our new proposal  by $\hat{\theta}\pm$$Z_{1-\frac{\alpha}{2}}\times$ S.E.$(\hat{\theta})$, where $Z_{1-\frac{\alpha}{2}}$ denoting the $1-\frac{\alpha}{2}$ quantile of the standard normal distribution, and S.E.$(\hat{\theta})$ is calculated by the inverse of the Fisher information matrix.  The results for this confidence interval were denoted by MLE(N) in the summary table. All these methods were implemented with
non-linear optimization techniques, thus we also summarized the number of converged cases (\emph{NOC}) in each scenario. For our method, the non-convergence was defined as a failure in optimization with negative hessian matrix or unsuccessful convergence based on the convergence indicator in the output of \emph{nlminb} function. For the Copas sensitivity analysis method, we followed the rule of \emph{Copas} function\cite{carpenter2009, carpenter2009r}.

Results of the simulation studies with ($\alpha_0,\alpha_1$)=(-2.18,0.2) were reported in Table 1 ($\rho=-0.4$) and Table 2 ($\rho=-0.8$), respectively. The datasets for Tables 1 and 2 had around 40\% unpublished studies on average. In almost all the simulated datasets, the log-likelihood was successfully maximized and then gave estimates. In all the scenarios, the REML had considerable biases. With $\rho=-0.8$ (Table 2), biases were larger, reflecting that $\rho=-0.8$ modeled a more selective publication process. Both the Copas method and our new proposal decreased biases and ours had the smallest biases in most scenarios. With larger number of studies (N=50 and 100), the maximum likelihood estimator tended to have very negligible biases,  and the confidence intervals based on the maximum likelihood theory (MLE(N)) had empirical coverage probabilities close to the nominal level of 95\%. With smaller number of the studies (N=15 and 25), our method successfully decreased biases. However, coverage probabilities might not be close to the nominal level and were not satisfactory when between-study heterogeneity was moderate or substantial ($\tau=0.15$ and 0.3). We demonstrated results of the simulation studies with ($\alpha_0,\alpha_1$)=(-1.24,0.16) in Tables S3 and S4, and that with  ($\alpha_0,\alpha_1$)=(-0.58,0.13) in Tables S5 and S6 in Web-appendix B. Findings were very similar to discussed above based on Tables 1 and 2.

\subsection{Modified confidence intervals}

To improve the coverage probabilities with smaller number of studies, we introduced two
 modifed confidence intervals. As suggested by Follmann and Proschan\cite{follmann1999}, an alternative confidence interval $\hat{\theta}\pm$$t_{(N-1);(1-\frac{\alpha}{2})}\times$ S.E.$(\hat{\theta})$ may be used, where $t_{(N-1);(1-\frac{\alpha}{2})}$ denoting the  $1-\frac{\alpha}{2}$ quantile of Student\emph{-t} distribution with $N-1$ degree of  freedom. Further modification was also considered by mimicking the idea of the improved estimator of variance which was proposed by Knapp and Hartung\cite{knapp2003}in the meta-regression context; we modified the standard error of  $\hat{\theta}$ by S.E.$(\hat{\theta})^\sharp$=Max(S.E.$(\hat{\theta})_{MLE}$, S.E.$(\hat{\theta})_{REML-KnHa}$) , where S.E.$(\hat{\theta})_{REML-KnHa}$ was the estimator of standard error with HK-adjusted REML method, reflecting our belief that the standard error of our proposal should be no less than the estimator of methods without considering the selcetive publication process for the  additional uncentainty of unpublished studies. That is, a more conservative confidence interval was calculated by  $\hat{\theta}\pm$$t_{(N-1);(1-\frac{\alpha}{2})}\times$S.E.$(\hat{\theta})^\sharp$. We applied these two confidence intervals to the simulated meta-analyses, results were also shown in Tables 1 and 2 and Tables S3-S6 in Web-appendix B, referring as MLE(T) and MLE($SE^\sharp$), respectively. We observed that these two confidence intervals MLE(T) and MLE($SE^\sharp$) gave certain improvement. In particular, MLE($SE^\sharp$) had much improved emprical coverage probabilities with \emph{N=15}.

\section{Re-analysis of two motivating examples with clinical trial registries}\label{sec5}

In this section, we applied our new inference procedure to the two case studies of the tiotropium study and clopidogrel study introduced in Section 2. To present data of unpublished studies as well as that of published studies, we propose to use a modified funnel-plot adding information of the planned sample size of unpublished studies with horizontal lines passing the y-axis at $\sqrt{n_i}$. The modified funnel-plots for tiotropium study and clopidogrel study were presented in Figures 2 and 3, respectively.  We could observe that for both of the two motivation cases, all the unpublished studies concentrated on the lower part of the funnel-plot. To the data, we applied our proposed method, as well as other related methods used in the simulation section, and the results for the two motivation cases were summarized in  Tables 3 and 4, respectively. 

As presented in Subsection 2.1, for the tiotropium study, we identified 24 published and 8 unpublished studies. By applying the linear mixed-effect model for the log-odds ratios with the REML method, the integrated odds ratio was obtanied as 0.768 with a two-tailed 95\% CI of [0.697,0.847] based on the normal approximation. The Knapp and Hartung modification gave a similar CI of [0.691, 0.854]. According to the sensitivity analysis by the Copas model, one can observe that as increasing the number of unpublished studies, the odds ratio increased up to 0.811, which was corresponding to the case of 13 unpublished studeis. The overall treatment effect was significantly different from unity even with 13 unpublished studies. Suppose we have 8 published studies behind which was  same with our findings from the clinical trial registries, Copas sensitivity analysis method gave the pooled odds rato as 0.803 with a 95\% CI of [0.717, 0.898]. Our proposed method gave the integrated estimate of 0.787 with two-tailed 95\% confidence intervals [0.710, 0.873] and [0.706, 0.878] based on the standard normal quantile and t-quantile, respectively.  In this case, the standard error from our method was larger than the standard error from REML with the Knapp and Hartung modification, so the modified confidence interval MLE($SE^\sharp$)  was same with MLE(T). To summary, all the methods provided statistically significant effect of tiotropium, indicating that selective publication might have not been much of an issue here. 

For the clopidogrel study which we introduced in Subsection 2.2, with only 3 unpublished studies identified from the clinical trial registries, the estimates with and without consideration of publication bias were considerably different. By applying the REML with 12 published studies, the estimator of the pooled odds ratio was 0.579 and with two-tailed 95\% CIs of [0.375, 0.892]  and [0.385, 0.871] based on the normal quantile and the Knapp and Hartung modification, respectively. Both of them supported the conclusion that the high maintenance-dose clopidogrel significantly reduced the incidence of MACE/MACCE in comparison to the standard-maintenance-dose clopidogrel.  However, the sensitivity analysis by the Copas model showed that as the number of unpublished studies increased up to 4, the statistical significance of the  High-maintenance-dose clopidogrel disappeared, and as high as 39\% change in estimates of odds ratio could be observed with 6 unpublished studies. Our proposed method gave the integrated estimate of 0.692, and a two-tailed 95\% CI based on the normal approximation was given as [0.496, 0.967], whose upper bound was very close to 1. The modified CIs introduced in Subsection 4.4 could be obtained as [0.476, 1.007] (MLE(T)) and [0.460, 1.041] (MLE($SE^\sharp$)), respectively. These results suggested that the significant effect of the high maintenance-dose clopidogrel might be marginal and should be interpreted with caution.

\section{Discussion}\label{sec6}

By utilizing the information on unpublished studies obtained from the clinical trial registries, we proposed a new inference procedure for the Copas selection model, which provides more objective evaluation of publication bias than the widely-used funnel plot and trim-and-fill method\cite{carpenter2009, schwarzer2010}. All the unknown parameters in the Copas selection model can be estimated from data with the proposed method. It resolved the issue of the sensitivity analysis approach that some unknown parameters had to be fixed in a certain range, and then gave a more insightful interpretation. Recently, Ning et al. \cite{ning2017}proposed a method to estimate all the unknown parameters. Whereas their method strongly relied on  an imputation based on funnel-plot symmetry, our method does not.

Here, we would like to address some issues in clinical trial registries.  Firstly, different clinical trial registries may give inconsistent information. A comparison of results between ClinicalTrials.gov and FDA database by Schwartz et al.\cite{schwartz2016}showed the planned sample size registered on the clinical trial registries were not always consistent with the final sample size for statistical analysis. Besides, as acknowledged by Fleminger et al.\cite{fleminger2018} the status of the studies might be inconsistent between EUCTR and ClinicalTrials.gov, the studies registered on  EUCTR which marked with ``completed'' may marked with ``onging'' in ClinicalTrials.gov. Secondly, no clinical trial registry could cover all the related studies\cite{glanville2014,tse2018,adam2018}. Thirdly, it is hard to integrate multiple clinical trial registries automatically since unique identifier of a study is not available among registries (e.g. World Health Organization's (WHO) International Clinical Trial Registry Platform (ICTRP)). Due to these issues of clinical trial registries, comprehensive search of multiple clinical trial registries was recommended in practice.

Clinical trial registries provide various information other than sample size. ClinicalTrials.gov's search result contains 26 items in the list and analysis result if the investigator submitted, ICTRP contains 40 items in the result file, and FDA database provide much more details about the registered clinical trials. As acknowledged by several cross-sectional studies, information such as funding type (government, industry or academic) and region may possibly play important roles in the publication status\cite{loder2018}. In addition, information of ongoing studies are also available, although we did not use them in this paper. How to take a further utilization of the clinical trial registry would deserve more attention.

\section*{Acknowledgments}
The last author's research was partly supported by Grant-in-Aid for Challenging Exploratory Research (16K12403) and for Scientific Research
(16H06299, 18H03208)  from the Ministry of Education, Science, Sports and Technology of Japan.

\newpage
\begin{figure}[t]
\centering\includegraphics[width=16cm, height=12cm]{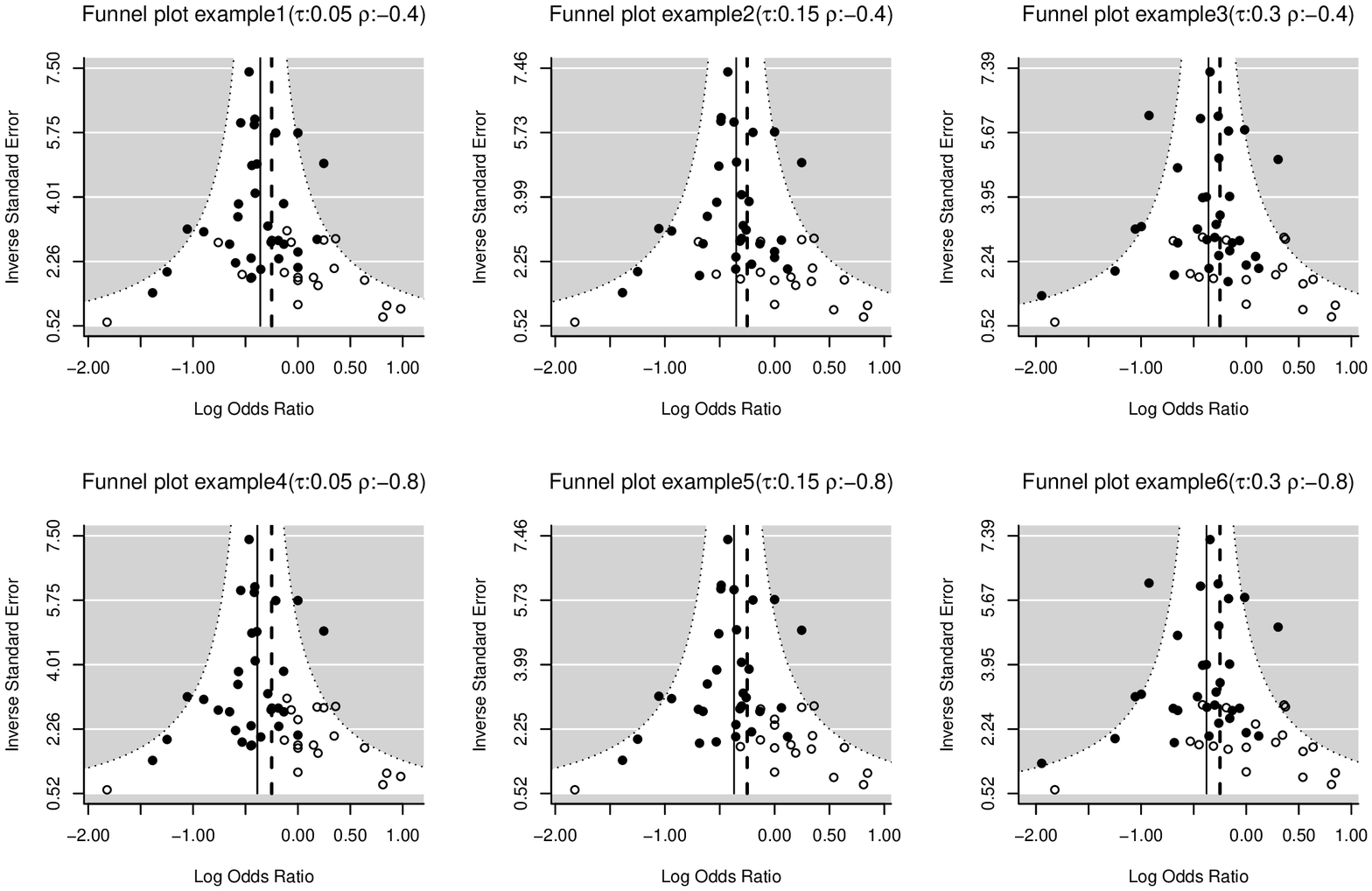}
\caption{\it{Funnel plots of randomly selected published (filled circles)/unpublished (open circles) studies in simulated datasets with $N+M=50$ under different settings of $\tau$ and $\rho$ }}
\label{fig1}
\end{figure}

\newpage
\begin{figure}[t]
\centering\includegraphics[width=16cm, height=12cm]{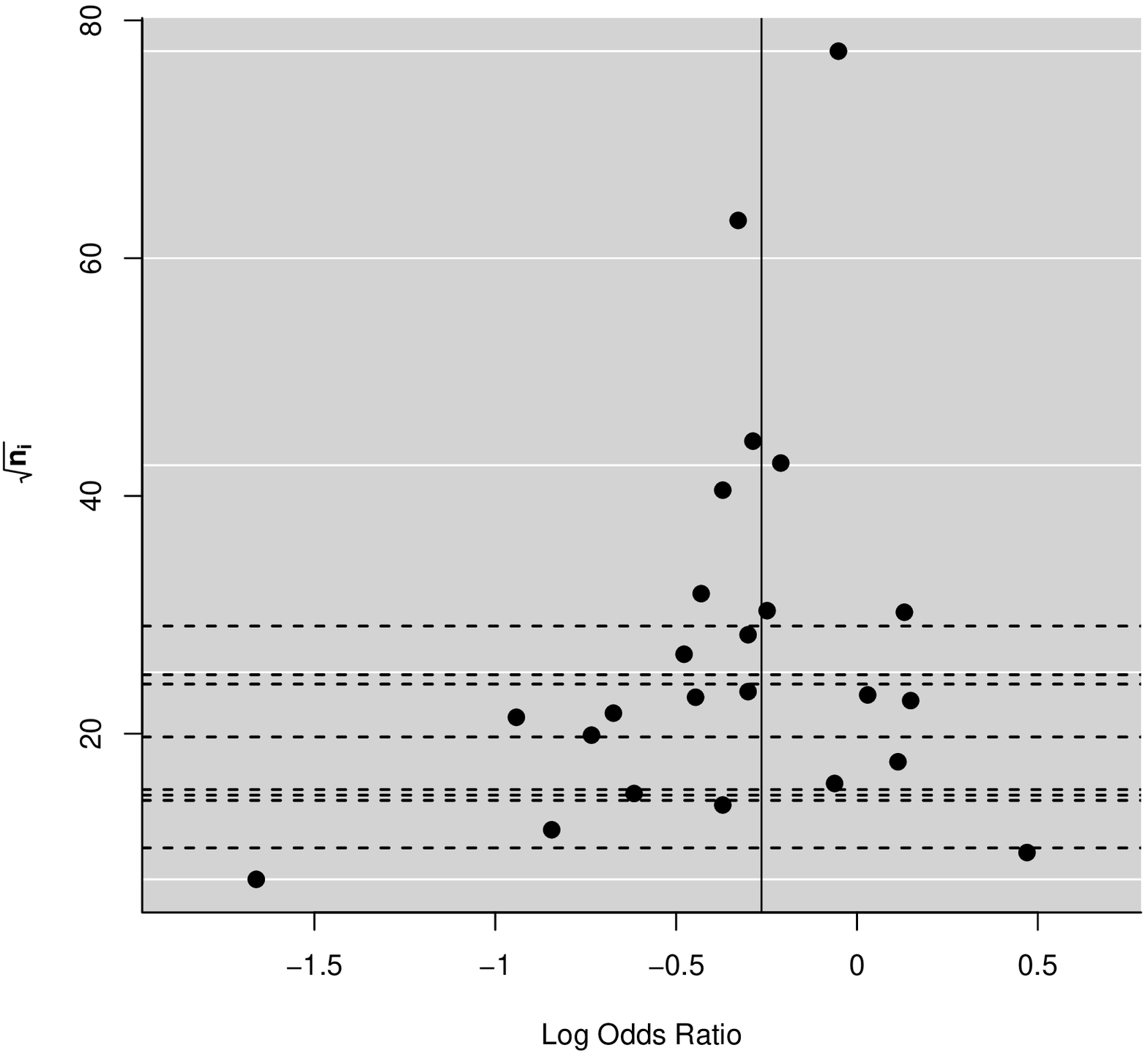}
\caption{\it{A funnel-plot for the tiotropium study being modified by adding information of the planned sample size of 8 unpublished studies with horizontal lines passing the y-axis at $\sqrt{n_i}$.}}
\label{fig 2}
\end{figure}

\newpage
\begin{figure}[t]
\centering\includegraphics[width=16cm, height=12cm]{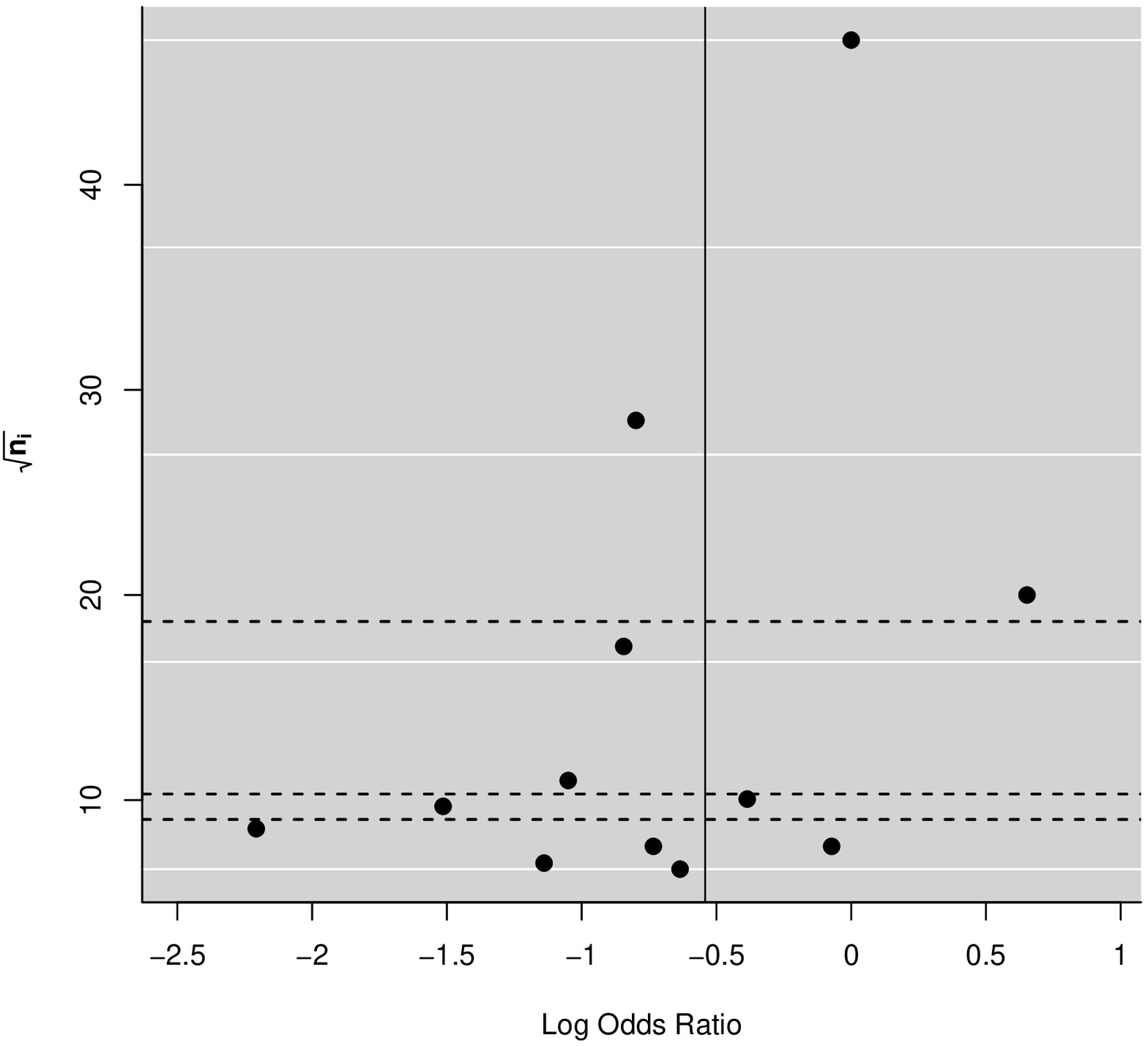}
\caption{\it{A funnel-plot for the clopidogrel study being modified by adding information of the planned sample size of 3 unpublished studies with horizontal lines passing the y-axis at $\sqrt{n_i}$.}}
\label{fig 3}
\end{figure}
\newpage

\begin{sidewaystable}
\centering
\captionsetup{justification=centering}

\caption{Simulation results for estimation of $\theta$ with $(\alpha_0,\alpha_1,\rho)$=(-2.18,0.2,-0.4) and $\tau=0.05, 0.15 \, or\, 0.30$ \protect \\ (40\% unpublished studies on average)}
\scalebox{0.75}[1]{%
\begin{tabular}{*{18}{c}}
\toprule
&&\multicolumn{4}{c}{$N+M=15$}&\multicolumn{4}{c}{$N+M=25$}&\multicolumn{4}{c}{$N+M=50$}&\multicolumn{4}{c}{$N+M=100$}\\
\cmidrule(lr){3-6}\cmidrule(lr){7-10}\cmidrule(lr){11-14}\cmidrule(lr){15-18}
Scenario&Method&AVE(SD)&CP&LOCI&NOC&AVE(SD)&CP&LOCI&NOC&AVE(SD)&CP&LOCI&NOC&AVE(SD)&CP&LOCI&NOC\\
\midrule
$\tau=0.05$
    & REML & -0.285(0.094) & 0.955 & 0.377 & 997 & -0.282(0.067) & 0.941 & 0.275 & 994 & -0.280(0.046) & 0.902 & 0.187 & 991 & -0.277(0.032) & 0.868 & 0.129 & 988 \\ 
    & REML.KnHa & -0.285(0.094) & 0.944 & 0.413 & 997 & -0.282(0.067) & 0.931 & 0.286 & 994 & -0.280(0.046) & 0.892 & 0.186 & 991 & -0.277(0.032) & 0.861 & 0.128 & 988 \\ 
    & Copas & -0.273(0.099) & 0.949 & 0.363 & 991 & -0.267(0.075) & 0.925 & 0.269 & 983 & -0.264(0.058) & 0.891 & 0.186 & 986 & -0.255(0.048) & 0.837 & 0.132 & 981 \\ 
    & MLE(N) & -0.257(0.112) & 0.904 & 0.368 & 968 & -0.251(0.077) & 0.930 & 0.289 & 995 & -0.249(0.054) & 0.935 & 0.202 & 998 & -0.248(0.035) & 0.955 & 0.140 & 999 \\ 
    & MLE(T) & -0.257(0.112) & 0.950 & 0.446 & 968 & -0.251(0.077) & 0.950 & 0.319 & 995 & -0.249(0.054) & 0.945 & 0.211 & 998 & -0.248(0.035) & 0.956 & 0.143 & 999 \\ 
    & MLE($SE^\sharp$) & -0.257(0.112) & 0.952 & 0.483 & 968 & -0.251(0.077) & 0.957 & 0.330 & 995 & -0.249(0.054) & 0.947 & 0.213 & 998 & -0.248(0.035) & 0.957 & 0.144 & 999 \\

 \midrule
\specialrule{0em}{5pt}{5pt}  

$\tau=0.15$    
    & REML & -0.286(0.112) & 0.920 & 0.420 & 999 & -0.288(0.079) & 0.913 & 0.310 & 993 & -0.286(0.056) & 0.893 & 0.216 & 994 & -0.283(0.04) & 0.854 & 0.153 & 997 \\ 
    & REML.KnHa& -0.286(0.112) & 0.932 & 0.481 & 999 & -0.288(0.079) & 0.928 & 0.333 & 993 & -0.286(0.056) & 0.903 & 0.223 & 994 & -0.283(0.04) & 0.852 & 0.155 & 997 \\ 
    & Copas & -0.268(0.128) & 0.873 & 0.392 & 984 & -0.271(0.091) & 0.879 & 0.299 & 986 & -0.265(0.072) & 0.862 & 0.215 & 993 & -0.256(0.059) & 0.825 & 0.156 & 996 \\ 
    & MLE(N) & -0.258(0.130) & 0.849 & 0.403 & 963 & -0.255(0.091) & 0.899 & 0.325 & 990 & -0.252(0.067) & 0.921 & 0.237 & 995 & -0.249(0.044) & 0.943 & 0.171 & 1000 \\ 
    & MLE(T) & -0.258(0.130) & 0.913 & 0.491 & 963 & -0.255(0.091) & 0.928 & 0.358 & 990 & -0.252(0.067) & 0.927 & 0.247 & 995 & -0.249(0.044) & 0.950 & 0.174 & 1000 \\ 
    & MLE($SE^\sharp$) & -0.258(0.13) & 0.938 & 0.541 & 963 & -0.255(0.091) & 0.942 & 0.374 & 990 & -0.252(0.067) & 0.932 & 0.250 & 995 & -0.249(0.044) & 0.950 & 0.175 & 1000 \\ 

  \midrule
\specialrule{0em}{5pt}{5pt} 
 
 $\tau=0.30$  
    & REML & -0.294(0.149) & 0.905 & 0.544 & 999 & -0.297(0.107) & 0.899 & 0.414 & 999 & -0.299(0.075) & 0.888 & 0.296 & 1000 & -0.295(0.054) & 0.867 & 0.210 & 1000 \\ 
    & REML.KnHa& -0.294(0.149) & 0.923 & 0.651 & 999 & -0.297(0.107) & 0.924 & 0.457 & 999 & -0.299(0.075) & 0.906 & 0.311 & 1000 & -0.295(0.054) & 0.881 & 0.216 & 1000 \\ 
    & Copas & -0.273(0.173) & 0.852 & 0.508 & 967 & -0.282(0.121) & 0.866 & 0.412 & 982 & -0.274(0.093) & 0.861 & 0.294 & 993 & -0.262(0.075) & 0.843 & 0.223 & 996 \\ 
    & MLE(N) & -0.262(0.168) & 0.827 & 0.529 & 947 & -0.261(0.126) & 0.876 & 0.442 & 984 & -0.253(0.092) & 0.915 & 0.334 & 994 & -0.249(0.062) & 0.948 & 0.244 & 999 \\ 
    & MLE(T) & -0.262(0.168) & 0.884 & 0.643 & 947 & -0.261(0.126) & 0.900 & 0.488 & 984 & -0.253(0.092) & 0.921 & 0.349 & 994 & -0.249(0.062) & 0.958 & 0.249 & 999 \\ 
    & MLE($SE^\sharp$) & -0.262(0.168) & 0.918 & 0.717 & 947 & -0.261(0.126) & 0.926 & 0.513 & 984 & -0.253(0.092) & 0.937 & 0.356 & 994 & -0.249(0.062) & 0.958 & 0.250 & 999 \\

   \midrule
&True&-0.250&-&-&-&-0.250&-&-&-&-0.250&-&-&-&-0.250&-&-&-\\
\bottomrule
\end{tabular}}
\begin{tablenotes}
\item N,number of total studies;AVE, mean value of estimates;SD,standard error of estimates; CP, 95\%confidence interval coverage probability; LOCI,length of confidence interval;NOC,number of converged cases.
\end{tablenotes}
\end{sidewaystable}

\begin{sidewaystable}
\centering
\captionsetup{justification=centering}
\caption{Simulation results for estimation of $\theta$ with $(\alpha_0,\alpha_1,\rho)$=(-2.18,0.2,-0.8) and $\tau=0.05, 0.15 \, or\, 0.30$\protect\\ (40\% unpublished studies on average)}
\scalebox{0.75}[1]{%
\begin{tabular}{*{18}{c}}
\toprule
&&\multicolumn{4}{c}{$N+M=15$}&\multicolumn{4}{c}{$N+M=25$}&\multicolumn{4}{c}{$N+M=50$}&\multicolumn{4}{c}{$N+M=100$}\\
\cmidrule(lr){3-6}\cmidrule(lr){7-10}\cmidrule(lr){11-14}\cmidrule(lr){15-18}
Scenario&Method&AVE(SD)&CP&LOCI&NOC&AVE(SD)&CP&LOCI&NOC&AVE(SD)&CP&LOCI&NOC&AVE(SD)&CP&LOCI&NOC\\
\midrule
$\tau=0.05$
  & REML & -0.320(0.097) & 0.913 & 0.376 & 998 & -0.313(0.068) & 0.881 & 0.272 & 992 & -0.307(0.046) & 0.789 & 0.186 & 992 & -0.304(0.033) & 0.644 & 0.128 & 983 \\ 
    & REML.KnHa & -0.320(0.097) & 0.917 & 0.404 & 998 & -0.313(0.068) & 0.874 & 0.275 & 992 & -0.307(0.046) & 0.769 & 0.182 & 992 & -0.304(0.033) & 0.615 & 0.125 & 983 \\ 
    & Copas & -0.291(0.105) & 0.906 & 0.365 & 982 & -0.276(0.090) & 0.862 & 0.267 & 969 & -0.249(0.081) & 0.773 & 0.190 & 933 & -0.220(0.066) & 0.664 & 0.138 & 929 \\ 
    & MLE(N) & -0.264(0.100) & 0.918 & 0.352 & 971 & -0.254(0.071) & 0.942 & 0.264 & 995 & -0.250(0.050) & 0.946 & 0.186 & 1000 & -0.249(0.033) & 0.956 & 0.130 & 1000 \\ 
    & MLE(T) & -0.264(0.100) & 0.968 & 0.426 & 971 & -0.254(0.071) & 0.957 & 0.291 & 995 & -0.250(0.050) & 0.954 & 0.194 & 1000 & -0.249(0.033) & 0.958 & 0.132 & 1000 \\ 
    & MLE($SE^\sharp$) & -0.264(0.100) & 0.973 & 0.466 & 971 & -0.254(0.071) & 0.962 & 0.307 & 995 & -0.250(0.050) & 0.965 & 0.200 & 1000 & -0.249(0.033) & 0.959 & 0.134 & 1000 \\ 
    
 \midrule
\specialrule{0em}{5pt}{5pt}  

$\tau=0.15$    
    & REML & -0.323(0.107) & 0.870 & 0.400 & 995 & -0.322(0.080) & 0.841 & 0.303 & 993 & -0.319(0.056) & 0.745 & 0.213 & 993 & -0.316(0.041) & 0.603 & 0.151 & 997 \\ 
    & REML.KnHa & -0.323(0.107) & 0.886 & 0.450 & 995 & -0.322(0.080) & 0.860 & 0.321 & 993 & -0.319(0.056) & 0.759 & 0.218 & 993 & -0.316(0.041) & 0.611 & 0.152 & 997 \\ 
    & Copas & -0.286(0.130) & 0.840 & 0.380 & 979 & -0.278(0.110) & 0.794 & 0.295 & 964 & -0.249(0.096) & 0.725 & 0.215 & 957 & -0.215(0.078) & 0.660 & 0.163 & 926 \\ 
    & MLE(N) & -0.266(0.116) & 0.888 & 0.369 & 973 & -0.259(0.084) & 0.909 & 0.293 & 991 & -0.254(0.061) & 0.924 & 0.216 & 996 & -0.249(0.041) & 0.937 & 0.157 & 998 \\ 
    & MLE(T) & -0.266(0.116) & 0.934 & 0.443 & 973 & -0.259(0.084) & 0.939 & 0.323 & 991 & -0.254(0.061) & 0.934 & 0.226 & 996 & -0.249(0.041) & 0.939 & 0.160 & 998 \\ 
    & MLE($SE^\sharp$) & -0.266(0.116) & 0.954 & 0.493 & 973 & -0.259(0.084) & 0.950 & 0.348 & 991 & -0.254(0.061) & 0.942 & 0.235 & 996 & -0.249(0.041) & 0.944 & 0.162 & 998 \\ 
    
  \midrule
\specialrule{0em}{5pt}{5pt} 
 
 $\tau=0.30$  
   & REML & -0.346(0.141) & 0.832 & 0.517 & 999 & -0.343(0.109) & 0.832 & 0.406 & 998 & -0.342(0.076) & 0.748 & 0.292 & 1000 & -0.339(0.054) & 0.604 & 0.208 & 1000 \\ 
    & REML.KnHa & -0.346(0.141) & 0.878 & 0.608 & 999 & -0.343(0.109) & 0.874 & 0.447 & 998 & -0.342(0.076) & 0.775 & 0.306 & 1000 & -0.339(0.054) & 0.613 & 0.213 & 1000 \\ 
    & Copas & -0.308(0.170) & 0.791 & 0.483 & 955 & -0.297(0.138) & 0.809 & 0.395 & 960 & -0.268(0.116) & 0.765 & 0.299 & 965 & -0.224(0.099) & 0.729 & 0.236 & 951 \\ 
    & MLE(N) & -0.285(0.164) & 0.822 & 0.484 & 942 & -0.270(0.120) & 0.875 & 0.406 & 982 & -0.259(0.085) & 0.917 & 0.314 & 985 & -0.251(0.058) & 0.943 & 0.225 & 986 \\ 
    & MLE(T) & -0.285(0.164) & 0.877 & 0.581 & 942 & -0.270(0.120) & 0.900 & 0.447 & 982 & -0.259(0.085) & 0.929 & 0.328 & 985 & -0.251(0.058) & 0.946 & 0.229 & 986 \\ 
    & MLE($SE^\sharp$) & -0.285(0.164) & 0.912 & 0.659 & 942 & -0.270(0.120) & 0.930 & 0.484 & 982 & -0.259(0.085) & 0.946 & 0.340 & 985 & -0.251(0.058) & 0.951 & 0.233 & 986 \\ 
   
   \midrule
&True&-0.250&-&-&-&-0.250&-&-&-&-0.250&-&-&-&-0.250&-&-&-\\
\bottomrule
\end{tabular}}
\begin{tablenotes}
\item N,number of total studies;AVE, mean value of estimates;SD,standard error of estimates; CP, 95\%confidence interval coverage probability; LOCI,length of confidence interval;NOC,number of converged cases.
\end{tablenotes}
\end{sidewaystable}

\begin{sidewaystable}
\centering
\captionsetup{justification=centering}
\caption{Summary of the statistical analysis for publication bias evaluation of tiotropium study:\\ \protect \# implies expected number of unpublished studies in the sensitivity analysis with the Copas selection model}
\begin{tabular}{ccccccc}
 \toprule
 Description & Method & \# & OR & 95\% CI&P-value  \\
 \midrule
No adjustment &  REML & - & 0.768&[0.697, 0.847]&$<$.000\\ 
&REML-KnHa&-&0.768&[0.691, 0.854]&$<$.000\\\\
Sensitivity analysis & Copas& 1 & 0.779 &[0.707, 0.858]&$<$.000\\
& & 5  &0.795 &[0.712, 0.886]&$<$.000\\
& & 8 &0.803 &[0.717, 0.898]&$<$.000\\
& & 11 &0.808 &[0.720, 0.907]&$<$.000\\
& & 13 &0.811 &[0.721, 0.912]&0.001\\ \\
Proposed &  MLE(N)&-&0.787&[0.710, 0.873]&$<$.000\\
&  MLE(T)&-&0.787&[0.706, 0.878]&$<$.000\\
\bottomrule
\end{tabular}

\end{sidewaystable}
\clearpage

\begin{sidewaystable}
\centering
\captionsetup{justification=centering}

\caption{Summary of the statistical analysis for publication bias evaluation of clopidogrel study:\\ \protect\# implies expected number of unpublished studies in the sensitivity analysis with the Copas selection model}
\begin{tabular}{ccccccc}
 \toprule
 Description & Method & \# & OR & 95\% CI&P-value  \\
 \midrule
No adjustment &  REML & - &0.579&[0.375, 0.892]&0.013\\ 
&REML-KnHa&-&0.579&[0.385, 0.871]&0.013\\\\
Sensitivity analysis & Copas
    & 2  &0.690 &[0.501, 0.952]&0.024\\
& & 3 &0.708 &[0.514, 0.975]&0.034\\
& & 4 &0.740 &[0.537, 1.020]&0.066\\ 
& & 6 &0.805 &[0.503, 1.288]&0.365\\
& & 8 &0.782 &[0.437, 1.397]&0.406\\\\
Proposed &  MLE(N)&-&0.692&[0.496, 0.967]&0.031\\
&  MLE(T)&-&0.692&[0.476, 1.007]&0.054\\
&  MLE($SE^\sharp$)&-&0.692&[0.460, 1.041]&0.073\\
\bottomrule
\end{tabular}
\end{sidewaystable}

\clearpage
\begin{center}
\Large 
Web-appendix to ``Using clinical trial registries to inform Copas selection model for publication bias in meta-analysis''
\end{center}

\vspace{3mm}
\begin{abstract}

In Web-appendix A, we provide the datasets of the tiotropium study and the clopidogrel study with information of published/unpublished studies. In Web-appendix B we present
the results of simulation studies with ($\alpha_0,\alpha_1$)=(-1.24,0.16) and (-0.58,0.13), respectively.
\end{abstract}

\appendix
\setcounter{table}{0}
\section*{Web-appendix A: Tables for the datasets of the tiotropium study and the clopidogrel study}%
In this appendix, we presented the datasets of the tiotropium study and the clopidogrel study.

In Table S1, the dataset of the tiotropium study was presented. Studies from No.1 to No.22 came from the original meta-analysis by Karner et al.\cite{karner20142} which we introduced in Subsection 2.1. By searching with multiple clinical trial registries, we further identified 10 studies with the same search terms: \emph{Tiotropium} or \emph{Spiriva} or \emph{HandiHaler} or \emph{Respimat} as the original meta-analysis. The identified 10 studies referred to the studies as from No.23 to No.32. Recall that $D_i$ is an indicator of publication status; $D_i=1$ if published and $D_i=0$ otherwise. Among these ten studies, study No.23 and No.24 reported estimated treatment effects in clinical trial registries. Therefore, they had the same indicator $D_i=1$ as the original study No.1 to No.22. The rest of them (from No.25 to No.32) were deemed as unpublished studies with the indicator $D_i=0$. The column of ``Status'' denotes the status of publication and registration. Studies marked with ``published $\&$ regsitered'' means we can identify these studies from both online literature database and clinical trial registries, while those marked with ``published'' or ``'registered' means they can only be identified from the online literature database or clinical trial registries.

In Table S2, the dataset of the clopidogrel study was presented. Similarly to the dataset of the tiotropium study, studies with $D_i=1$ (from No.1 to No.12) came from the original meta-analysis by Chen et al. \cite{chen20132}. The studies with $D_i=0$ were identified from the clinical trial registries.
\renewcommand{\thetable}{S\arabic{table}}
\begin{sidewaystable}
\centering
\caption{Tiotropium dataset}
\begin{tabular}{rlrrrrrrrrl}
  \toprule
  &&\multicolumn{2}{c}{Tiotropium}&\multicolumn{2}{c}{Placebo}\\
  \cmidrule(lr){3-4}\cmidrule(lr){5-6}
  
 No.& Study &Events & Total & Events & Total & $n_i$ & $logOR_i$ & $s_i$ & $D_i$ & Status \\ 
  \midrule
  1 & Bateman 2010a & 685 & 1989 & 842 & 2002 & 3991 & -0.33 & 0.07 & 1 & published \& registered \\ 
  2 & Bateman 2010b & 495 & 1337 & 288 & 653 & 1990 & -0.29 & 0.10 & 1 & published \& registered \\ 
  3 & Beeh 2006 & 180 & 1236 & 80 & 403 & 1639 & -0.37 & 0.15 &1 & published \& registered \\ 
  4 & Brusasco 2003 & 129 & 402 & 156 & 400 & 802 & -0.30 & 0.15 &1 & published \\ 
  5 & Casaburi 2002 & 198 & 550 & 156 & 371 & 921 & -0.25 & 0.14 & 1 & published \\ 
  6 & Chan 2007 & 268 & 608 & 125 & 305 & 913 & 0.13 & 0.14 & 1 & published \& registered \\ 
  7 & Cooper 2010 & 112 & 260 & 102 & 259 & 519 & 0.15 & 0.18 & 1 & published \& registered \\ 
  8 & Covelli 2005 & 9 & 100 & 12 & 96 & 196 & -0.37 & 0.47 & 1 & published \& registered \\ 
  9 & Dusser 2006 & 213 & 500 & 272 & 510 & 1010 & -0.43 & 0.13 & 1 & published \& registered \\ 
  10 & Freeman 2007 & 19 & 200 & 35 & 195 & 395 & -0.73 & 0.30 & 1 & published \& registered \\ 
  11 & Johansson 2008 & 2 & 107 & 4 & 117 & 224 & -0.62 & 0.88 & 1 & published \& registered \\ 
  12 & Magnussen 2008 & 13 & 228 & 26 & 244 & 472 & -0.67 & 0.35 & 1 & published \& registered \\ 
  13 & Moita 2008 & 6.00 & 147 & 6 & 164 & 311 & 0.11 & 0.59 & 1 & published \& registered \\ 
  14 & NCT00144326 & 11 & 123 & 12 & 127 & 250 & -0.06 & 0.44 & 1 & registered \\ 
  15 & Niewoehner 2005 & 255 & 914 & 296 & 915 & 1829 & -0.21 & 0.10 & 1 & published \& registered \\ 
  16 & Powrie 2007 & 30 & 69 & 47 & 73 & 142 & -0.84 & 0.34 &1 & published \& registered \\ 
  17 & Sun 2007 & 0 & 30 & 2 & 30 & 60 & -1.66 & 1.57 & 1 & published \\ 
  18 & Tashkin 2008 & 2001 & 2987 & 2049 & 3006 & 5993 & -0.05 & 0.06 & 1& published \& registered \\ 
  19 & Tonnel 2008 & 101 & 266 & 130 & 288 & 554 & -0.30 & 0.17 & 1 & published \& registered \\ 
  20 & Trooster 2011 & 11 & 238 & 24 & 219 & 457 & -0.94 & 0.38 & 1 & published \& registered \\ 
  21 & Verkinde 2006 & 10 & 46 & 8 & 54 & 100 & 0.47 & 0.52 & 1 & published \\ 
  22 & Voshaar 2008 & 43 & 360 & 21 & 181 & 541 & 0.03 & 0.28 & 1 & published \& registered \\ 
  23 & NCT01202188 & 85 & 480 & 60 & 232 & 712 & -0.48 & 0.19 & 1& registered \\ 
  24 & NCT00929110 & 80 & 266 & 107 & 266 & 532 & -0.45 & 0.18 &1& registered \\ 
  25 & NCT00668772 &  &  &  &  & 207 &  &  & 0 & registered \\ 
  26 & NCT00662740 &  &  &  &  & 220 &  &  &  0 & registered \\ 
  27 & NCT00157235 &  &  &  &  & 234 &  &  &  0 & registered \\ 
  28 & NCT00274521 &  &  &  &  & 108 &  &  &  0 & registered \\ 
  29 & NCT02172287 &  &  &  &  & 623 &  &  &  0 & registered \\ 
  30 & NCT02173691 &  &  &  &  & 584 &  &  &  0 & registered \\ 
  31 & 2007-001946-42-ES &  &  &  &  & 844 &  &  &  0&  registered\\ 
  32 & 2005-000650-79 &  &  &  &  & 389 &  &  & 0 & registered \\ 
   \bottomrule
\end{tabular}
\end{sidewaystable}

\begin{sidewaystable}
\centering
\caption{Clopidogrel dataset}
\begin{tabular}{rlrrrrrrrrl}
  \toprule
  &&\multicolumn{2}{c}{High Dose}&\multicolumn{2}{c}{Standard Dose}\\
  \cmidrule(lr){3-4}\cmidrule(lr){5-6}
  
 No.& Study &Events & Total & Events & Total & $n_i$ &$ logOR_i$ &$ s_i $& $D_i$ & Status \\ 
  \midrule
1 & Aradi 2012 & 1 & 36 & 8 & 38 & 74 & -2.23 & 1.09 & 1 & published \& registered \\ 
  2 & DOUBLE 2010 & 0 & 24 & 1 & 24 & 48 & -1.14 & 1.66 & 1 & published \\ 
  3 & EFFICIENT 2011 & 2 & 47 & 8 & 47 & 94 & -1.53 & 0.82 & 1 & published \& registered \\ 
  4 & GRAVITAS 2011 & 25 & 1109 & 25 & 1105 & 2214 & -0.00 & 0.29 & 1 & published \& registered \\ 
  5 & Gremmel 2011 & 1 & 21 & 2 & 23 & 44 & -0.64 & 1.26 & 1 & published \\ 
  6 & Han 2009 & 4 & 403 & 9 & 410 & 813 & -0.81 & 0.61 & 1 & published \\ 
  7 & Ren LH 2012 & 6 & 46 & 10 & 55 & 101 & -0.39 & 0.56 & 1 & published \\ 
  8 & Roghani 2011 & 4 & 205 & 2 & 195 & 400 & 0.65 & 0.87 & 1 & published \\ 
  9 & Tousek 2011 & 1 & 30 & 2 & 30 & 60 & -0.73 & 1.25 & 1 & published \\ 
  10 & VASP-02 2008 & 0 & 58 & 1 & 62 & 120 & -1.05 & 1.64 & 1 & published \& registered \\ 
  11 & von Beckerath 2007 & 1 & 31 & 1 & 29 & 60 & -0.07 & 1.44 & 1 & published \\ 
  12 & Wang 2011 & 14 & 150 & 30 & 156 & 306 & -0.84 & 0.35 & 1 & published \\ 
  13 & NCT01069302 &  &  &  &  & 106 &  &  & 0 & registered \\ 
  14 & NCT01371058 &  &  &  &  & 350 &  &  & 0 & registered \\ 
  15 & NCT01102439 &  &  &  &  & 82 &  &  & 0 & registered \\
  \bottomrule
\end{tabular}
\end{sidewaystable}
 \clearpage

 \section*{Web-appendix B: Tables for simulation results with additional settings}%

In this appendix, we presented the results of simulation studies with ($\alpha_0,\alpha_1$)=(-1.24,0.16) (Table S3 and Table S4) and (-0.58,0.13) (Table S5 and Table S6),  which resulted in the 27\% and 19\%
unpublished studies on average, respectively. The findings of these two settings were similar with the results in Tables 1 and 2.  We observed that our method had the smallest biases in all the scenarios in comparison
to the REML and the Copas sensitivity analysis method. With larger number of studies (N=50 and 100), the confidence intervals of our method based on the normal approximation had emperical coverage probabilities
close to the nominal level of 95\%. With smaller number of studies (N=15 and 25), the modified confidence intervals could be useful to improve the coverage probabilities of our method when between-study variance
was considerable.

\begin{sidewaystable}
\centering
\captionsetup{justification=centering}

\caption{Simulation results for estimation of $\theta$ with $(\alpha_0,\alpha_1,\rho)$=(-1.24,0.16,-0.4) and $\tau=0.05, 0.15 \, or\, 0.30$  \protect \\(27\% unpublished studies on average)}
\scalebox{0.75}[1]{%
\begin{tabular}{*{18}{c}}
\toprule
&&\multicolumn{4}{c}{$N+M=15$}&\multicolumn{4}{c}{$N+M=25$}&\multicolumn{4}{c}{$N+M=50$}&\multicolumn{4}{c}{$N+M=100$}\\
\cmidrule(lr){3-6}\cmidrule(lr){7-10}\cmidrule(lr){11-14}\cmidrule(lr){15-18}
Scenario&Method&AVE(SD)&CP&LOCI&NOC&AVE(SD)&CP&LOCI&NOC&AVE(SD)&CP&LOCI&NOC&AVE(SD)&CP&LOCI&NOC\\
\midrule
$\tau=0.05$
   & REML & -0.280(0.083) & 0.957 & 0.353 & 991 & -0.276(0.063) & 0.942 & 0.262 & 997 & -0.275(0.044) & 0.915 & 0.179 & 988 & -0.272(0.031) & 0.891 & 0.124 & 989 \\ 
    & REML.KnHa & -0.280(0.083) & 0.954 & 0.370 & 991 & -0.276(0.063) & 0.945 & 0.265 & 997 & -0.275(0.044) & 0.901 & 0.177 & 988 & -0.272(0.031) & 0.882 & 0.122 & 989 \\ 
    & Copas & -0.270(0.092) & 0.944 & 0.344 & 994 & -0.268(0.066) & 0.943 & 0.257 & 992 & -0.263(0.054) & 0.896 & 0.178 & 991 & -0.257(0.040) & 0.870 & 0.126 & 989 \\ 
    & MLE(N) & -0.261(0.096) & 0.927 & 0.351 & 972 & -0.253(0.072) & 0.932 & 0.272 & 994 & -0.250(0.053) & 0.930 & 0.191 & 999 & -0.249(0.034) & 0.952 & 0.135 & 999 \\ 
    & MLE(T) & -0.261(0.096) & 0.954 & 0.402 & 972 & -0.253(0.072) & 0.948 & 0.293 & 994 & -0.250(0.053) & 0.935 & 0.198 & 999 & -0.249(0.034) & 0.954 & 0.138 & 999 \\ 
    & MLE($SE^\sharp$) & -0.261(0.096) & 0.965 & 0.422 & 972 & -0.253(0.072) & 0.953 & 0.300 & 994 & -0.250(0.053) & 0.937 & 0.200 & 999 & -0.249(0.034) & 0.954 & 0.138 & 999 \\ 
     
 \midrule
\specialrule{0em}{5pt}{5pt}  

$\tau=0.15$    
   & REML & -0.287(0.099) & 0.920 & 0.390 & 994 & -0.280(0.074) & 0.919 & 0.293 & 990 & -0.281(0.053) & 0.900 & 0.205 & 993 & -0.278(0.038) & 0.878 & 0.145 & 997 \\ 
    & REML.KnHa & -0.287(0.099) & 0.925 & 0.426 & 994 & -0.280(0.074) & 0.923 & 0.307 & 990 & -0.281(0.053) & 0.908 & 0.209 & 993 & -0.278(0.038) & 0.880 & 0.146 & 997 \\ 
    & Copas & -0.276(0.109) & 0.898 & 0.371 & 992 & -0.267(0.085) & 0.894 & 0.285 & 993 & -0.263(0.070) & 0.861 & 0.203 & 996 & -0.259(0.049) & 0.863 & 0.148 & 998 \\ 
    & MLE(N) & -0.264(0.120) & 0.871 & 0.379 & 973 & -0.255(0.085) & 0.913 & 0.304 & 995 & -0.252(0.064) & 0.912 & 0.225 & 1000 & -0.250(0.042) & 0.941 & 0.163 & 1000 \\ 
    & MLE(T) & -0.264(0.120) & 0.916 & 0.435 & 973 & -0.255(0.085) & 0.931 & 0.328 & 995 & -0.252(0.064) & 0.921 & 0.233 & 1000 & -0.250(0.042) & 0.944 & 0.165 & 1000 \\ 
    & MLE($SE^\sharp$) & -0.264(0.120) & 0.934 & 0.466 & 973 & -0.255(0.085) & 0.946 & 0.340 & 995 & -0.252(0.064) & 0.924 & 0.235 & 1000 & -0.250(0.042) & 0.945 & 0.166 & 1000 \\

  \midrule
\specialrule{0em}{5pt}{5pt} 
 
 $\tau=0.30$  
   & REML & -0.286(0.134) & 0.917 & 0.505 & 998 & -0.293(0.103) & 0.906 & 0.389 & 999 & -0.291(0.072) & 0.895 & 0.277 & 1000 & -0.288(0.051) & 0.882 & 0.197 & 1000 \\ 
    & REML.KnHa & -0.286(0.134) & 0.940 & 0.570 & 998 & -0.293(0.103) & 0.917 & 0.417 & 999 & -0.291(0.072) & 0.912 & 0.287 & 1000 & -0.288(0.051) & 0.887 & 0.200 & 1000 \\ 
    & Copas & -0.271(0.150) & 0.872 & 0.479 & 985 & -0.275(0.119) & 0.868 & 0.378 & 996 & -0.271(0.089) & 0.864 & 0.274 & 997 & -0.265(0.065) & 0.866 & 0.201 & 998 \\ 
    & MLE(N) & -0.261(0.153) & 0.859 & 0.503 & 962 & -0.264(0.119) & 0.885 & 0.413 & 985 & -0.254(0.087) & 0.930 & 0.313 & 995 & -0.250(0.059) & 0.942 & 0.228 & 999 \\ 
    & MLE(T) & -0.261(0.153) & 0.905 & 0.577 & 962 & -0.264(0.119) & 0.905 & 0.445 & 985 & -0.254(0.087) & 0.943 & 0.324 & 995 & -0.250(0.059) & 0.943 & 0.232 & 999 \\ 
    & MLE($SE^\sharp$) & -0.261(0.153) & 0.928 & 0.619 & 962 & -0.264(0.119) & 0.922 & 0.460 & 985 & -0.254(0.087) & 0.948 & 0.327 & 995 & -0.250(0.059) & 0.946 & 0.232 & 999 \\ 
   
   \midrule
&True&-0.250&-&-&-&-0.250&-&-&-&-0.250&-&-&-&-0.250&-&-&-\\
\bottomrule
\end{tabular}}
\begin{tablenotes}
\item N,number of total studies;AVE, mean value of estimates;SD,standard error of estimates; CP, 95\%confidence interval coverage probability; LOCI,length of confidence interval;NOC,number of converged cases.
\end{tablenotes}
\end{sidewaystable}

\begin{sidewaystable}
\centering
\captionsetup{justification=centering}

\caption{Simulation results for estimation of $\theta$ with $(\alpha_0,\alpha_1,\rho)$=(-1.24,0.16,-0.8) and $\tau=0.05, 0.15 \, or\, 0.30$ \protect \\(27\% unpublished studies on average)}
\scalebox{0.75}[1]{%
\begin{tabular}{*{18}{c}}
\toprule
&&\multicolumn{4}{c}{$N+M=15$}&\multicolumn{4}{c}{$N+M=25$}&\multicolumn{4}{c}{$N+M=50$}&\multicolumn{4}{c}{$N+M=100$}\\
\cmidrule(lr){3-6}\cmidrule(lr){7-10}\cmidrule(lr){11-14}\cmidrule(lr){15-18}
Scenario&Method&AVE(SD)&CP&LOCI&NOC&AVE(SD)&CP&LOCI&NOC&AVE(SD)&CP&LOCI&NOC&AVE(SD)&CP&LOCI&NOC\\
\midrule
$\tau=0.05$
  & REML & -0.304(0.084) & 0.932 & 0.345 & 989 & -0.302(0.063) & 0.901 & 0.258 & 991 & -0.297(0.043) & 0.840 & 0.177 & 985 & -0.295(0.031) & 0.712 & 0.123 & 982 \\ 
    & REML.KnHa & -0.304(0.084) & 0.912 & 0.348 & 989 & -0.302(0.063) & 0.881 & 0.251 & 991 & -0.297(0.043) & 0.811 & 0.168 & 985 & -0.295(0.031) & 0.669 & 0.116 & 982 \\ 
    & Copas & -0.281(0.097) & 0.912 & 0.336 & 985 & -0.274(0.081) & 0.888 & 0.254 & 983 & -0.259(0.069) & 0.824 & 0.179 & 958 & -0.234(0.058) & 0.728 & 0.130 & 969 \\ 
    & MLE(N) & -0.262(0.088) & 0.933 & 0.331 & 981 & -0.255(0.068) & 0.931 & 0.252 & 994 & -0.251(0.047) & 0.939 & 0.177 & 998 & -0.249(0.031) & 0.952 & 0.123 & 1000 \\ 
    & MLE(T) & -0.262(0.088) & 0.958 & 0.379 & 981 & -0.255(0.068) & 0.952 & 0.271 & 994 & -0.251(0.047) & 0.945 & 0.184 & 998 & -0.249(0.031) & 0.956 & 0.125 & 1000 \\ 
    & MLE($SE^\sharp$) & -0.262(0.088) & 0.963 & 0.398 & 981 & -0.255(0.068) & 0.955 & 0.280 & 994 & -0.251(0.047) & 0.950 & 0.186 & 998 & -0.249(0.031) & 0.956 & 0.126 & 1000 \\ 
    
 \midrule
\specialrule{0em}{5pt}{5pt}  

$\tau=0.15$    
    & REML & -0.317(0.100) & 0.902 & 0.380 & 994 & -0.310(0.074) & 0.870 & 0.286 & 989 & -0.307(0.052) & 0.800 & 0.201 & 992 & -0.305(0.038) & 0.664 & 0.142 & 994 \\ 
    & REML.KnHa & -0.317(0.100) & 0.899 & 0.404 & 994 & -0.310(0.074) & 0.869 & 0.292 & 989 & -0.307(0.052) & 0.793 & 0.200 & 992 & -0.305(0.038) & 0.658 & 0.140 & 994 \\ 
    & Copas & -0.290(0.118) & 0.874 & 0.366 & 989 & -0.274(0.100) & 0.832 & 0.280 & 988 & -0.256(0.088) & 0.760 & 0.201 & 978 & -0.231(0.070) & 0.690 & 0.151 & 964 \\ 
    & MLE(N) & -0.274(0.116) & 0.884 & 0.361 & 968 & -0.259(0.080) & 0.906 & 0.282 & 995 & -0.254(0.057) & 0.918 & 0.207 & 993 & -0.250(0.038) & 0.938 & 0.148 & 999 \\ 
    & MLE(T) & -0.274(0.116) & 0.928 & 0.414 & 968 & -0.259(0.080) & 0.925 & 0.303 & 995 & -0.254(0.057) & 0.928 & 0.214 & 993 & -0.250(0.038) & 0.941 & 0.151 & 999 \\ 
    & MLE($SE^\sharp$) & -0.274(0.116) & 0.939 & 0.444 & 968 & -0.259(0.080) & 0.937 & 0.318 & 995 & -0.254(0.057) & 0.938 & 0.218 & 993 & -0.250(0.038) & 0.943 & 0.152 & 999 \\ 
    
  \midrule
\specialrule{0em}{5pt}{5pt} 
 
 $\tau=0.30$  
   & REML & -0.322(0.135) & 0.878 & 0.482 & 1000 & -0.328(0.099) & 0.862 & 0.377 & 999 & -0.327(0.070) & 0.797 & 0.272 & 1000 & -0.324(0.051) & 0.668 & 0.193 & 1000 \\ 
    & REML.KnHa & -0.322(0.135) & 0.899 & 0.537 & 1000 & -0.328(0.099) & 0.877 & 0.401 & 999 & -0.327(0.070) & 0.808 & 0.279 & 1000 & -0.324(0.051) & 0.675 & 0.194 & 1000 \\ 
    & Copas & -0.290(0.156) & 0.824 & 0.459 & 968 & -0.288(0.130) & 0.804 & 0.369 & 985 & -0.270(0.106) & 0.779 & 0.275 & 991 & -0.239(0.091) & 0.726 & 0.213 & 990 \\ 
    & MLE(N) & -0.272(0.149) & 0.851 & 0.464 & 975 & -0.270(0.109) & 0.882 & 0.384 & 986 & -0.259(0.079) & 0.931 & 0.294 & 994 & -0.251(0.054) & 0.946 & 0.211 & 991 \\ 
    & MLE(T) & -0.272(0.149) & 0.893 & 0.531 & 975 & -0.270(0.109) & 0.906 & 0.413 & 986 & -0.259(0.079) & 0.939 & 0.304 & 994 & -0.251(0.054) & 0.950 & 0.215 & 991 \\ 
    & MLE($SE^\sharp$) & -0.272(0.149) & 0.917 & 0.578 & 975 & -0.270(0.109) & 0.924 & 0.433 & 986 & -0.259(0.079) & 0.947 & 0.311 & 994 & -0.251(0.054) & 0.952 & 0.217 & 991 \\ 

   \midrule
  &True&-0.250&-&-&-&-0.250&-&-&-&-0.250&-&-&-&-0.250&-&-&-\\
\bottomrule
\end{tabular}}
\begin{tablenotes}
\item N,number of total studies;AVE, mean value of estimates;SD,standard error of estimates; CP, 95\%confidence interval coverage probability; LOCI,length of confidence interval;NOC,number of converged cases.
\end{tablenotes}
\end{sidewaystable}

\begin{sidewaystable}
\centering
\captionsetup{justification=centering}

\caption{Simulation results for estimation of $\theta$ with $(\alpha_0,\alpha_1,\rho)$=(-0.58,0.13,-0.4) and $\tau=0.05, 0.15 \, or\, 0.30$ \protect \\(19\% unpublished studies on average)}
\scalebox{0.75}[1]{%
\begin{tabular}{*{18}{c}}
\toprule
&&\multicolumn{4}{c}{$N+M=15$}&\multicolumn{4}{c}{$N+M=25$}&\multicolumn{4}{c}{$N+M=50$}&\multicolumn{4}{c}{$N+M=100$}\\
\cmidrule(lr){3-6}\cmidrule(lr){7-10}\cmidrule(lr){11-14}\cmidrule(lr){15-18}
Scenario&Method&AVE(SD)&CP&LOCI&NOC&AVE(SD)&CP&LOCI&NOC&AVE(SD)&CP&LOCI&NOC&AVE(SD)&CP&LOCI&NOC\\
\midrule
$\tau=0.05$
   & REML & -0.274(0.082) & 0.960 & 0.346 & 994 & -0.268(0.063) & 0.941 & 0.254 & 989 & -0.271(0.043) & 0.932 & 0.176 & 987 & -0.268(0.031) & 0.911 & 0.122 & 985 \\ 
    & REML.KnHa & -0.274(0.082) & 0.945 & 0.357 & 994 & -0.268(0.063) & 0.934 & 0.254 & 989 & -0.271(0.043) & 0.922 & 0.172 & 987 & -0.268(0.031) & 0.903 & 0.119 & 985 \\ 
    & Copas & -0.265(0.088) & 0.951 & 0.337 & 996 & -0.261(0.067) & 0.928 & 0.250 & 997 & -0.261(0.051) & 0.908 & 0.174 & 992 & -0.258(0.036) & 0.898 & 0.122 & 995 \\ 
    & MLE(N) & -0.261(0.095) & 0.928 & 0.343 & 969 & -0.253(0.071) & 0.929 & 0.260 & 996 & -0.253(0.050) & 0.934 & 0.186 & 998 & -0.250(0.033) & 0.944 & 0.132 & 1000 \\ 
    & MLE(T) & -0.261(0.095) & 0.954 & 0.386 & 969 & -0.253(0.071) & 0.940 & 0.277 & 996 & -0.253(0.050) & 0.943 & 0.192 & 998 & -0.250(0.033) & 0.951 & 0.134 & 1000 \\ 
    & MLE($SE^\sharp$) & -0.261(0.095) & 0.961 & 0.403 & 969 & -0.253(0.071) & 0.942 & 0.283 & 996 & -0.253(0.050) & 0.944 & 0.194 & 998 & -0.250(0.033) & 0.951 & 0.134 & 1000 \\ 
 \midrule
\specialrule{0em}{5pt}{5pt}  

$\tau=0.15$    
  & REML & -0.274(0.096) & 0.940 & 0.378 & 995 & -0.269(0.074) & 0.929 & 0.286 & 986 & -0.275(0.052) & 0.912 & 0.200 & 991 & -0.272(0.037) & 0.904 & 0.142 & 997 \\ 
    & REML.KnHa & -0.274(0.096) & 0.932 & 0.406 & 995 & -0.269(0.074) & 0.933 & 0.295 & 986 & -0.275(0.052) & 0.917 & 0.203 & 991 & -0.272(0.037) & 0.904 & 0.142 & 997 \\ 
    & Copas & -0.264(0.103) & 0.916 & 0.363 & 995 & -0.258(0.082) & 0.905 & 0.278 & 993 & -0.263(0.064) & 0.883 & 0.203 & 999 & -0.260(0.044) & 0.892 & 0.143 & 996 \\ 
    & MLE(N) & -0.256(0.108) & 0.905 & 0.371 & 959 & -0.251(0.082) & 0.898 & 0.294 & 994 & -0.255(0.061) & 0.917 & 0.216 & 999 & -0.251(0.041) & 0.937 & 0.159 & 1000 \\ 
    & MLE(T) & -0.256(0.108) & 0.935 & 0.418 & 959 & -0.251(0.082) & 0.922 & 0.314 & 994 & -0.255(0.061) & 0.925 & 0.223 & 999 & -0.251(0.041) & 0.938 & 0.162 & 1000 \\ 
    & MLE($SE^\sharp$) & -0.256(0.108) & 0.949 & 0.442 & 959 & -0.251(0.082) & 0.931 & 0.323 & 994 & -0.255(0.061) & 0.927 & 0.225 & 999 & -0.251(0.041) & 0.938 & 0.162 & 1000 \\ 

  \midrule
\specialrule{0em}{5pt}{5pt} 
 
 $\tau=0.30$  
   & REML & -0.280(0.125) & 0.917 & 0.480 & 997 & -0.284(0.100) & 0.914 & 0.375 & 999 & -0.283(0.069) & 0.909 & 0.268 & 1000 & -0.280(0.050) & 0.896 & 0.191 & 1000 \\ 
    & REML.KnHa & -0.280(0.125) & 0.927 & 0.531 & 997 & -0.284(0.100) & 0.924 & 0.398 & 999 & -0.283(0.069) & 0.919 & 0.277 & 1000 & -0.280(0.050) & 0.898 & 0.193 & 1000 \\ 
    & Copas & -0.264(0.143) & 0.880 & 0.460 & 989 & -0.272(0.109) & 0.881 & 0.366 & 998 & -0.269(0.082) & 0.880 & 0.266 & 998 & -0.264(0.059) & 0.879 & 0.193 & 1000 \\ 
    & MLE(N) & -0.261(0.142) & 0.878 & 0.472 & 961 & -0.265(0.113) & 0.889 & 0.390 & 982 & -0.257(0.083) & 0.926 & 0.297 & 996 & -0.252(0.058) & 0.931 & 0.220 & 1000 \\ 
    & MLE(T) & -0.261(0.142) & 0.904 & 0.531 & 961 & -0.265(0.113) & 0.908 & 0.416 & 982 & -0.257(0.083) & 0.932 & 0.306 & 996 & -0.252(0.058) & 0.936 & 0.223 & 1000 \\ 
    & MLE($SE^\sharp$) & -0.261(0.142) & 0.922 & 0.564 & 961 & -0.265(0.113) & 0.916 & 0.428 & 982 & -0.257(0.083) & 0.937 & 0.309 & 996 & -0.252(0.058) & 0.937 & 0.223 & 1000 \\ 
     
   \midrule
&True&-0.250&-&-&-&-0.250&-&-&-&-0.250&-&-&-&-0.250&-&-&-\\
\bottomrule
\end{tabular}}
\begin{tablenotes}
\item
 N,number of total studies;AVE, mean value of estimates;SD,standard error of estimates; CP, 95\%confidence interval coverage probability; LOCI,length of confidence interval;NOC,number of converged cases.
\end{tablenotes}
\end{sidewaystable}

\begin{sidewaystable}
\centering
\captionsetup{justification=centering}

\caption{Simulation results for estimation of $\theta$ with  $(\alpha_0,\alpha_1,\rho)$=(-0.58,0.13,-0.8) and $\tau=0.05, 0.15 \, or\, 0.30$ \protect \\(19\% unpublished studies on average)}
\scalebox{0.75}[1]{%
\begin{tabular}{*{18}{c}}
\toprule
&&\multicolumn{4}{c}{$N+M=15$}&\multicolumn{4}{c}{$N+M=25$}&\multicolumn{4}{c}{$N+M=50$}&\multicolumn{4}{c}{$N+M=100$}\\
\cmidrule(lr){3-6}\cmidrule(lr){7-10}\cmidrule(lr){11-14}\cmidrule(lr){15-18}
Scenario&Method&AVE(SD)&CP&LOCI&NOC&AVE(SD)&CP&LOCI&NOC&AVE(SD)&CP&LOCI&NOC&AVE(SD)&CP&LOCI&NOC\\
\midrule
$\tau=0.05$
  & REML & -0.297(0.083) & 0.939 & 0.340 & 989 & -0.289(0.058) & 0.929 & 0.253 & 990 & -0.288(0.042) & 0.870 & 0.173 & 985 & -0.286(0.030) & 0.791 & 0.120 & 980 \\ 
    & REML.KnHa & -0.297(0.083) & 0.917 & 0.332 & 989 & -0.289(0.058) & 0.917 & 0.240 & 990 & -0.288(0.042) & 0.841 & 0.162 & 985 & -0.286(0.030) & 0.750 & 0.112 & 980 \\ 
    & Copas & -0.281(0.090) & 0.932 & 0.333 & 995 & -0.271(0.072) & 0.913 & 0.250 & 989 & -0.264(0.061) & 0.854 & 0.174 & 974 & -0.248(0.052) & 0.774 & 0.123 & 962 \\ 
    & MLE(N) & -0.267(0.090) & 0.932 & 0.327 & 976 & -0.255(0.066) & 0.941 & 0.249 & 995 & -0.253(0.047) & 0.936 & 0.173 & 999 & -0.250(0.031) & 0.956 & 0.120 & 1000 \\ 
    & MLE(T) & -0.267(0.090) & 0.958 & 0.368 & 976 & -0.255(0.066) & 0.958 & 0.266 & 995 & -0.253(0.047) & 0.941 & 0.178 & 999 & -0.250(0.031) & 0.958 & 0.122 & 1000 \\ 
    & MLE($SE^\sharp$) & -0.267(0.090) & 0.966 & 0.382 & 976 & -0.255(0.066) & 0.962 & 0.272 & 995 & -0.253(0.047) & 0.944 & 0.180 & 999 & -0.250(0.031) & 0.958 & 0.123 & 1000 \\ 

 \midrule
\specialrule{0em}{5pt}{5pt}  

$\tau=0.15$    
& REML & -0.295(0.097) & 0.913 & 0.368 & 988 & -0.293(0.072) & 0.902 & 0.279 & 987 & -0.296(0.051) & 0.839 & 0.195 & 989 & -0.294(0.037) & 0.756 & 0.138 & 992 \\ 
    & REML.KnHa & -0.295(0.097) & 0.915 & 0.381 & 988 & -0.293(0.072) & 0.899 & 0.278 & 987 & -0.296(0.051) & 0.834 & 0.192 & 989 & -0.294(0.037) & 0.733 & 0.134 & 992 \\ 
    & Copas & -0.276(0.109) & 0.877 & 0.359 & 988 & -0.267(0.094) & 0.849 & 0.272 & 986 & -0.263(0.078) & 0.802 & 0.195 & 990 & -0.248(0.061) & 0.754 & 0.143 & 973 \\ 
    & MLE(N) & -0.265(0.105) & 0.883 & 0.352 & 963 & -0.256(0.081) & 0.904 & 0.277 & 993 & -0.257(0.057) & 0.916 & 0.200 & 1000 & -0.251(0.038) & 0.932 & 0.144 & 1000 \\ 
    & MLE(T) & -0.265(0.105) & 0.915 & 0.396 & 963 & -0.256(0.081) & 0.930 & 0.296 & 993 & -0.257(0.057) & 0.924 & 0.206 & 1000 & -0.251(0.038) & 0.936 & 0.146 & 1000 \\ 
    & MLE($SE^\sharp$) & -0.265(0.105) & 0.929 & 0.417 & 963 & -0.256(0.081) & 0.936 & 0.305 & 993 & -0.257(0.057) & 0.932 & 0.209 & 1000 & -0.251(0.038) & 0.940 & 0.146 & 1000 \\ 
    
  \midrule
\specialrule{0em}{5pt}{5pt} 
 
 $\tau=0.30$  
   & REML & -0.310(0.129) & 0.887 & 0.470 & 994 & -0.312(0.096) & 0.885 & 0.362 & 998 & -0.312(0.068) & 0.838 & 0.262 & 1000 & -0.309(0.049) & 0.761 & 0.186 & 1000 \\ 
    & REML.KnHa & -0.310(0.129) & 0.901 & 0.513 & 994 & -0.312(0.096) & 0.891 & 0.379 & 998 & -0.312(0.068) & 0.840 & 0.266 & 1000 & -0.309(0.049) & 0.756 & 0.186 & 1000 \\ 
    & Copas & -0.284(0.149) & 0.847 & 0.450 & 982 & -0.286(0.119) & 0.851 & 0.356 & 994 & -0.273(0.097) & 0.803 & 0.261 & 996 & -0.256(0.077) & 0.779 & 0.195 & 994 \\ 
    & MLE(N) & -0.278(0.145) & 0.848 & 0.459 & 961 & -0.272(0.108) & 0.887 & 0.375 & 986 & -0.261(0.077) & 0.920 & 0.277 & 990 & -0.252(0.052) & 0.942 & 0.201 & 994 \\ 
    & MLE(T) & -0.278(0.145) & 0.889 & 0.517 & 961 & -0.272(0.108) & 0.901 & 0.401 & 986 & -0.261(0.077) & 0.932 & 0.285 & 990 & -0.252(0.052) & 0.950 & 0.204 & 994 \\ 
    & MLE($SE^\sharp$) & -0.278(0.145) & 0.911 & 0.550 & 961 & -0.272(0.108) & 0.916 & 0.414 & 986 & -0.261(0.077) & 0.936 & 0.289 & 990 & -0.252(0.052) & 0.950 & 0.205 & 994 \\ 
   
   \midrule
  &True&-0.250&-&-&-&-0.250&-&-&-&-0.250&-&-&-&-0.250&-&-&-\\
\bottomrule
\end{tabular}}
\begin{tablenotes}
\item
 N,number of total studies;AVE, mean value of estimates;SD,standard error of estimates; CP, 95\%confidence interval coverage probability; LOCI,length of confidence interval;NOC,number of converged cases.
\end{tablenotes}
\end{sidewaystable}

\end{document}